\documentclass[12pt]{iopart}
\usepackage{url}
\usepackage{times}
\usepackage{graphicx}
\usepackage{subfigure}

\begin{document}

\title{
A gain-coefficient switched Alexandrite laser
}
\author{Chris~J.~Lee$^{1,2}$, Peter~J.~M.~van~der~Slot$^{1}$, and Klaus-J.~Boller$^{1}$}
\address{$^{1}$Laser Physics and Nonlinear Optics group, MESA+ Research Institute for Nanotechnology, University of Twente, P.O.~Box~217, 7500~AE, Enschede, The Netherlands}
\address{$^{2}$FOM Institute DIFFER, Edisonbaan~14, 3439~MN, Nieuwegein, The Netherlands
}
\ead{c.j.lee@differ.nl}

\begin{abstract}
We report on a gain-coefficient switched Alexandrite laser. An electro-optic modulator is used to switch between high and low gain states by making use of the polarization dependent gain of Alexandrite. In gain-coefficient switched mode, the laser produces 85~ns pulses with a pulse energy of 240~mJ at a repetition rate of 5~Hz.
\end{abstract}

\submitto{\JPD}
\maketitle

\section{Introduction}
\label{Introduction}
Applications, such as photoacoustic mammography~\cite{Piras:2010ie}, lidar~\cite{Roy:2008p2385,Bruneau:1994p15112} and space applications~\cite{Schall:2002wz} require lasers that have high pulse energy. For photoacoustic applications, \emph{Q}-switched laser pulses are used to generate ultrasound waves comprising a wide range of frequency components. The acoustically acquired image resolution is limited by the highest acoustic frequency generated by the laser pulse. Optimum resolution requires that the ultrasound waves are initiated by a laser pulse that has a duration less than 100~ns (FWHM)~\cite{Piras:2010ie}.  In addition, the penetration depth of the laser light and the image contrast depend on the optical absorption of the tissue components, such that near IR lasers, e.g. Alexandrite lasers, have received increased attention~\cite{Kruger:2010gb}. An important consideration for lasers in industrial and medical applications is simplicity in design by minimizing the number of cavity components, allowing for robust and stable instruments. For \emph{Q}-switched lasers based on low gain materials, such as Alexandrite, this is of interest also from a fundamental point of view, because additional cavity components introduce additional loss, making the laser less efficient.   

\begin{figure}[!t]
\centering
\subfigure[]{\includegraphics[width=3in]{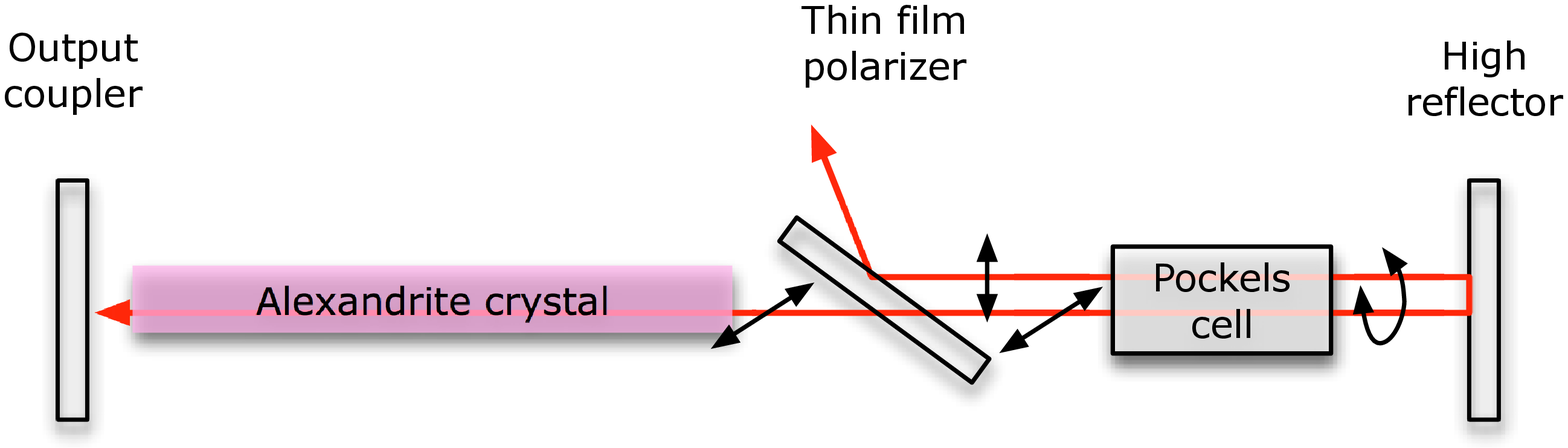}}\\
\subfigure[]{\includegraphics[width=3in]{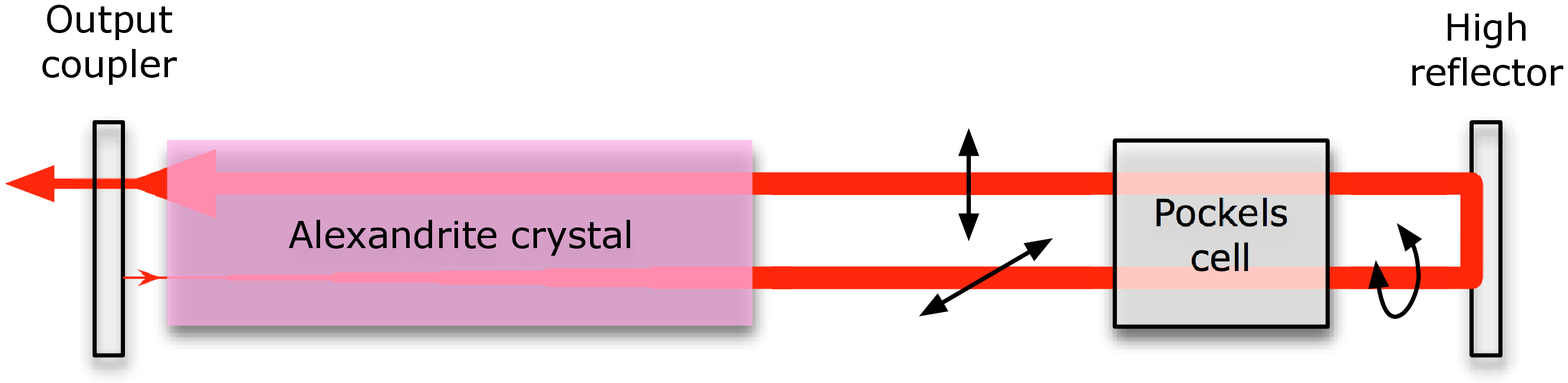}}
\caption{\emph{Q}-switching by loss modulation (a). The Pockels cell, combined with the thin film polarizer selects high and low loss states. Gain-coefficient switched operation (b). The Pockels cell rotates the polarization every roundtrip, coupling each polarization mode to the high and low gain cross sections of the laser gain. In this way, the Pockels cell increases the threshold of the laser, preventing laser action before the Pockels cell is switched off.}
\label{Q-switching}
\end{figure}

\emph{Q}-switching is achieved through loss modulation. For instance, a Pockels cell is placed in the optical cavity. In the hold-off state, the Pockels cell acts as a quarter wave plate, and, after two passes, the polarization state of the light is rotated by 90$^{\circ}$. Adding a polarizing beamsplitter then rejects the rotated light from the cavity, introducing large losses (see Fig.~\ref{Q-switching}(a)). When out of the hold-off state, the Pockels cell has no effect on the polarization of the light, reducing the losses of the cavity and allowing the laser to reach threshold. \emph{Q}-switching is an efficient way of producing nanosecond, high energy pulses because the switching device--usually a Pockels cell or acousto-optic modulator--is able to switch the cavity from closed to open in a time that is much shorter than the upper state lifetime of the gain medium.

However, losses that are inevitably associated with intracavity elements, increase the build-up time and reduce the output power of the oscillator. Although intracavity elements are optimized to be low loss, a low gain material is particularly sensitive to the accumulated effect of these losses. This is because it takes many round trips---with losses acquired on a per-pass basis---before the optical energy has grown by, typically, 20 orders of magnitude from spontaneously emitted photons and shows up as a macroscopic pulse. If losses delay this build-up process to an extent that becomes comparable with the lifetime of the upper laser level, a significant fraction of the energy stored in the inversion is lost, which reduces the output pulse energy. In cases like this, it is beneficial to consider pulse generating schemes that eliminate as many intracavity optics as possible.

Gain switching is an alternative that is, in general, considered to be less favourable for nanosecond pulse generation. In addition to requiring that the gain is switched faster than the life time of the upper state, it is also required that the gain is switched faster than the build up time of the laser system. Thus, although Alexandrite has an upper state life time of 1.54~ms~\cite{Verdeyen:wa}, a flash lamp pulse with a duration of 200~$\mu$s results in a burst of laser pulses consisting of weak nanosecond pulses emitted over a timespan that can extend into the millisecond range (an effect that is observed in many laser gain materials). To obtain a single nanosecond pulse, one would preferentially pump a gain switched laser with a \emph{Q}-switched laser to rapidly switch the gain (see e.g.,~\cite{Lee:2002uy}).

In this paper, we demonstrate that nanosecond pulses with high energy can be generated from what we call gain-coefficient switching. In all lasers, the amplification is a product of the population inversion and the coupling between the population and the light field, i.e., the Einstein $B$ gain-coefficient. Conventional gain switching relies on a rapid ramp of the population inversion, while keeping the gain-coefficient constant. However, in gain-coefficient switching, the population inversion need not be ramped rapidly, instead, the gain-coefficient is switched. We demonstrate gain-coefficient switching with an Alexandrite laser, where the Einstein $B$ gain-coefficient for stimulated emission is switched from a lower to a higher value. The observation of such effects date back to work from Walling \emph{et al.}, who obtained $Q$-switching and tuning without an intracavity polarizer~\cite{Walling:1980uk}. However, gain-coefficient switching is unlike the standard gain switching or Q-switching because both the inversion and cavity losses remain unchanged upon switching. Alexandrite is strongly birefringent and has a strongly polarization dependent gain coefficient---an order of magnitude more gain along one axis of birefringence compared to the other~\cite{koechner2006solid}. We use this to modulate the laser gain via a Pockels cell to switch the coupling between the light in the optical cavity and the population inversion via its Einstein $B$ coefficient. This allows the flash lamp to provide a large population inversion over a time interval that is much longer than the build up time of the cavity, while the Pockels cell prevents light, spontaneously emitted into the mode of the cavity, from coupling efficiently to the inversion. After the flash lamp pulse is complete, and the population inversion is built up, the Pockels cell is switched off, allowing the single polarization mode with the highest B coefficient to efficiently build up, generating a single nanosecond duration gain switched pulse.

\section{Setup}
\label{exp setup}
The laser is based around an Alexandrite crystal that was doped at 0.15\% Cr, 115~mm long, and 7~mm in diameter (Northrup Grumman). The end facets were polished at a 2$^{\circ}$ angle and anti-reflection coated at 755~nm. The laser crystal was placed in one the foci of an elliptical flash lamp housing (Quantel SF 511-07), oriented such that one of the axes of birefringence was parallel with the optical table. The laser crystal was excited by two Xenon flash lamps that were supplied by a capacitor bank that stored up to 281~J of energy.

The laser cavity consisted of two plane mirrors, separated by 31~cm (limited by the physical size of the laser crystal housing and the Pockels cell mount). The available output couplers had reflectivities of 90, 80, 70, and 60\%. We found that a 60\% reflective output coupler provided the shortest pulse duration, while the output energy for all output couplers was limited by the damage threshold of the optical coatings (15~J/cm$^{2}$). All results presented below were taken with the 60\% output coupler. Gain-coefficient switched operation was achieved by placing a Pockels cell between the laser crystal housing and the high reflector. The Pockels cell operation sequence was triggered by the flash lamp discharge control signal. At the same time that the capacitor bank discharge is triggered, a voltage of 1100~V is applied to the Pockels cell. The cell then behaves like a quarter wave plate, which is the hold-off mode. After a delay of 250~$\mu$s, the Pockels cell voltage is switched off with a fall time of 15~ns, coupling a single polarization mode to the highest B coefficient and allowing laser oscillation. The Pockels cell remains in this state until the next flash lamp discharge trigger signal. Careful examination of the time trace with the Pockels cell in hold-off mode revealed that the laser was unable to reach threshold even at the maximum flash lamp power.

\section{Results}
\label{exp results}
\begin{figure}[!t]
\centering
\includegraphics[width=3in]{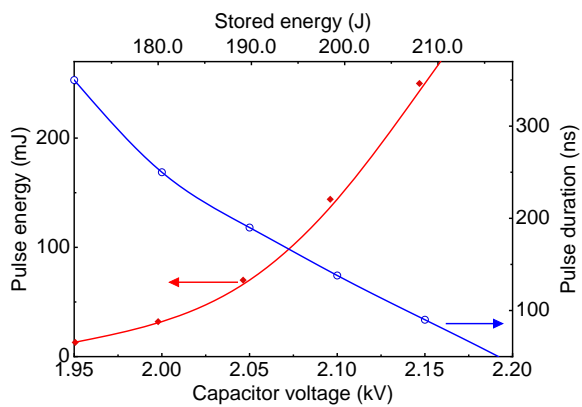}
\caption{Pulse energy (filled red diamonds) and pulse duration (FWHM) (open blue circles) as a function of energy stored in the capacitor bank. Lines are a guide to the eye}
\label{pulseEnergy}
\end{figure}

Fig.~\ref{pulseEnergy} shows the pulse energy and pulse duration as a function of energy stored in the energy of the capacitor bank. The laser threshold was found to be 170~J (1.95~kV). The pulse energy reaches a maximum value of 240~mJ at a stored energy of 208~J, limited by the damage threshold of the output coupler's optical coating. 

\begin{figure}[!t]
\centering
\includegraphics[width=3in]{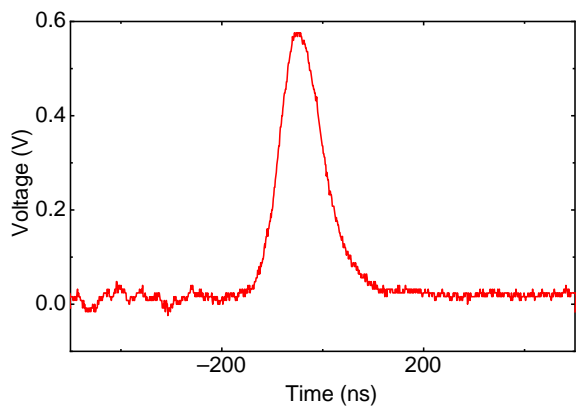}
\caption{Gain-coefficient switched pulse, obtained at flash lamp voltage of 2.15~kV.}
\label{pulse envelope}
\end{figure}

The full width half maximum (FWHM) pulse duration, measured by a photodiode with a rise-time of 1~ns (Thorlabs DET 10A), reduces from 350~ns to 85~ns as the flash lamp energy is increased (Fig.~\ref{pulseEnergy}). The delay between the Pockels cell closing and the centre of the laser pulse was measured to be 2.9~$\mu$s for a capacitor voltage of 2.15~kV. A typical example of the pulse envelope, for a capacitor charging voltage of 2.15~kV, is shown in Fig.~\ref{pulse envelope}. 

\begin{figure}[!t]
\centering
\includegraphics[width=3in]{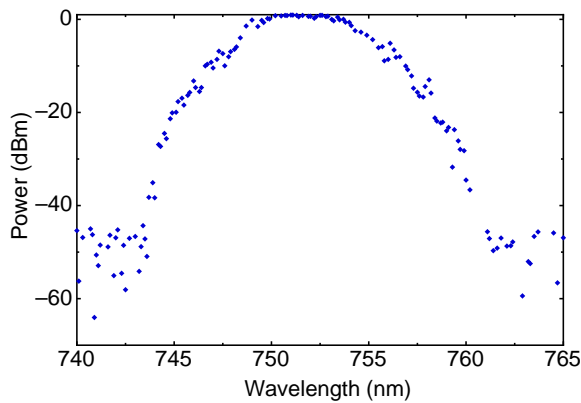}
\caption{Average output spectrum.}
\label{spectrum}
\end{figure}

The spectrum of the laser is shown in Fig.~\ref{spectrum}. The spectrum was acquired by averaging $\sim$100 laser shots using an optical spectrum analyser (Ando AQ6317, spectral resolution 0.05~nm), showing that the laser operates at a central wavelength of 753~nm, and has an average spectral bandwidth of 5~nm (FWHM). The mirror coatings, and the crystal anti-reflection coatings were chosen to optimize the laser for operation at wavelengths longer than 730~nm, where laser performance is also enhanced by high crystal temperatures (we estimate that the laser crystal temperature was about 100$^{\circ}$C)~\cite{Verdeyen:wa}.

\section{Discussion}
\label{sec:discussion}
A qualitative understanding of how rapidly switching the polarization of the light fields in the cavity can be used to gain switch the laser can be obtained by examining the threshold gain of the laser. At threshold, where gain equals loss, one obtains a threshold gain of~\cite{Verdeyen:wa}
\begin{equation}
\label{eq:threshold gain}
g_{th} = \frac{-1}{2}\log(R_{oc}T_{oc}) + \alpha_{rt}
\end{equation}
where $R_{oc}$ and $T_{oc}$ are the reflectivity and transmissivity of the output coupler, and $\alpha_{rt}$ is the lumped additional loss (e.g., all losses except those due to output coupling) per roundtrip of the optical cavity.  

The polarization of the light in the cavity can be described by two orthogonal, linearly polarized light fields with an arbitrary phase difference between the two. We consider two orthogonally polarized light fields, aligned to the high (horizontal) and low (vertical) gain axes of the Alexandrite crystal. Furthermore, since the small signal gain for the vertically polarized light field is an order of magnitude less than that of the horizontally polarized light field, it is considered that the gain of the vertically polarized light field is negligible. Under these conditions, laser oscillation is restricted to just the horizontally polarized light field. By introducing the Pockels cell in the hold-off state into the cavity, reducing the average roundtrip gain to below the laser threshold, without, however, changing the roundtrip loss. Under these conditions, the threshold gain for the horizontal polarization mode increases to
\begin{eqnarray}
\label{eq:threshold gain modified}
g'_{th} &=& \frac{-1}{2}\log(R_{oc}^{2}T_{oc}^{2}) + 2\alpha_{RT}\\
g_{th}/g'_{th} &=& 2
\end{eqnarray}
where $g'_{th}$ is the threshold gain for the laser when the Pockels cell is in hold-off mode and the light only experiences gain every second roundtrip. The single pass gain of the horizontal polarization needs to be twice the single pass losses as there is only gain every second roundtrip, while the losses are not changed. For the laser described above, the threshold energy is 170~J (see Fig.~\ref{pulseEnergy}), indicating that the Pockels cell should successfully hold-off up to pump energies of 340~J, considerably higher than the maximum stored energy in the capacitor bank.

A simplified rate equation model was used to confirm this conclusion. In this model, the maximum population inversion was fixed, the ground state is considered to be the lower laser level, and the excited states were considered to be identical except for the orientation of their dipole moments. At the beginning of the simulation, a set fraction of the population is instantly transferred to the excited states. Population transfer between the two excited states is neglected since the difference in gain coefficient indicates that population transfer between excited states is very slow. The ratio of the initial population in the two excited states was varied between 1 and 10. The change in population and growth in the light field intensities are then numerically integrated in time steps of 0.1~ns, while the photon life time in the optical cavity was set to 5~ns. The light intensity was subjected to fractional losses of 0.01 due to intracavity elements and 0.4 due to output coupling. For each excited state population ratio, the output fluence of the laser was calculated twice: once with Pockel's cell in hold-off mode and once with the Pockels cell turned off. 

\begin{figure}[!t]
\centering
\includegraphics[width=3in]{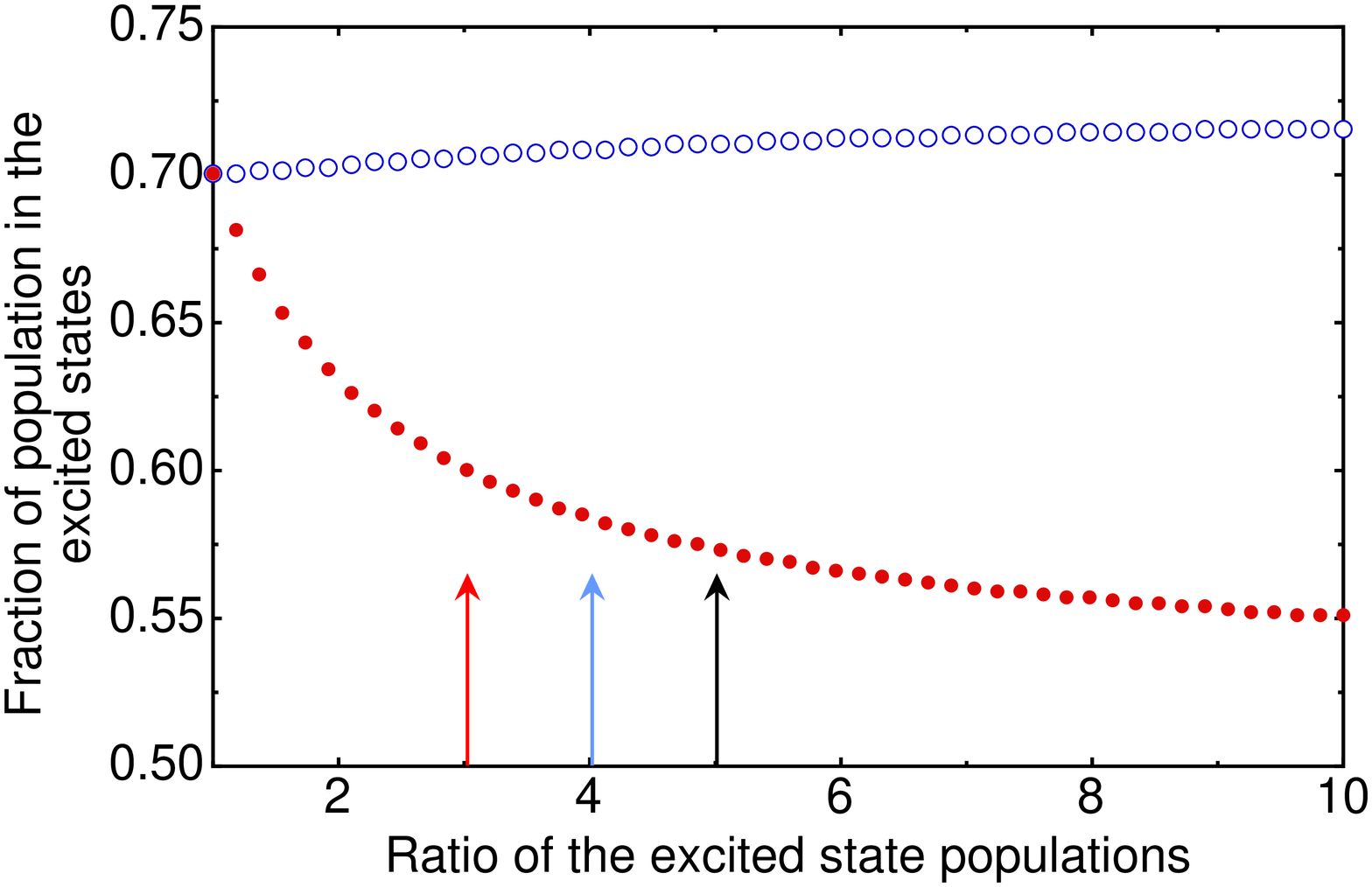}
\caption{Fraction of the total population in both excited states required to achieve threshold when the Pockels cell is passive (red closed circles) and when the Pockels cell is in hold-off mode (blue open circles) for different ratios of population between excited states. The arrows indicate the gain ratio's for Ti:sapphire (red), Nd:YLF (light blue), Ruby (black), }
\label{threshold}
\end{figure}

The fraction of population in the excited states required to reach threshold as a function of ratio of the initial population in the excited states is shown in Fig.~\ref{threshold}. The red closed circles show the threshold when the Pockels cell is inactive and the blue open circles show the threshold for when the Pockels cell is in hold-off mode. The difference in threshold for the two operating regimes increases as the ratio increases, as expected. However, the increase in threshold above the ideal (0.5) is a factor of 3.4, indicating that the ratio in the two populations does not have to be as large as it is for Alexandrite. Indeed, a useful threshold difference is obtained, even for population ratios as small as 2. The observed gain ratios for Ti:sapphire~\cite{Verdeyen:wa} (red arrow), Nd:YLF~\cite{Ryan:1992vt} (light blue arrow) and Ruby~\cite{Verdeyen:wa} (black arrow) are also shown in Fig.~\ref{threshold}, indicating that this technique may be of use for other laser gain materials as well.   

\section{Conclusions}
\label{sec: conclusions}
In conclusion, we have presented a gain-coefficient switched Alexandrite laser, emitting up to 240~mJ with a pulse duration of 85~ns and a repetition rate of 5~Hz. The system design takes advantage of the polarization dependent gain of Alexandrite to initiate a gain switched pulse through a Pockels cell, operating in quarter wave mode, but without introducing a polarization dependent loss. Instead, the polarization rotation introduced by the Pockels cell reduces the coupling between the light in the laser cavity and the Einstein $B$ coefficient of the laser gain medium, effectively preventing the laser from oscillating when in hold-off mode. A rate equation model indicates that this mode of operation may be applicable to other laser gain media, such as Ti:sapphire and Nd:YLF.
\section*{Acknowledgment}
The financial support of the Agentschap NL Innovation Oriented Research Programmes Photonic Devices under the HYMPACT Project (IPD083374) is acknowledged.

\section*{References}

\end{document}